\documentclass[lettersize,journal]{IEEEtran}
\usepackage{amsmath,amsfonts}
\usepackage{algorithmic}
\usepackage{algorithm}
\usepackage{array}
\usepackage[caption=false,font=normalsize,labelfont=sf,textfont=sf]{subfig}
\usepackage{textcomp}
\usepackage{stfloats}
\usepackage{url}
\usepackage{verbatim}
\usepackage{graphicx}
\usepackage{cite}
\usepackage{xcolor}
\usepackage{tabularx}
\usepackage{array}
\usepackage{colortbl}
\usepackage{tabularray}
\usepackage{wrapfig} %
\newtheorem{definition}{Definition}
\newcolumntype{C}[1]{>{\centering\arraybackslash}p{#1}}

\definecolor{darkgreen}{rgb}{0.0, 0.7, 0.1}
\definecolor{darkred}{rgb}{0.9, 0.0, 0.0}
\definecolor{darkblue}{rgb}{0.05, 0.2, 0.9}

\usepackage{fancyhdr}

\begin{document}

\title{Generative AI for RF Sensing in IoT systems}

\author{Li Wang, Chao Zhang, Qiyang Zhao, Hang Zou,\\
Samson Lasaulce, Giuseppe Valenzise, Zhuo He, and Merouane Debbah
\thanks{\textit{Li Wang and Chao Zhang are with the Central South University, Changsha, China, and Khalifa University, UAE.
Qiyang Zhao and Hang Zou are with Technology Innovation Institute, UAE.
Samson Lasaulce is with Khalifa University, UAE and CRAN, Nancy, France.
Giuseppe Vlenzise is with CNRS, France.
Zhuo He is with Michigan Technological University.
Merouane Debbah is with Khalifa University, UAE.}}
\thanks{\textit{(Corresponding authors: Chao Zhang(chao.zhang@csu.edu.cn), Zhuo He(zhuoh@mtu.edu))}}}



\maketitle

\begin{abstract}
The development of wireless sensing technologies, using signals such as Wi-Fi, infrared, and RF to gather environmental data, has significantly advanced within Internet of Things (IoT) systems. Among these, Radio Frequency (RF) sensing stands out for its cost-effective and non-intrusive monitoring of human activities and environmental changes. However, traditional RF sensing methods face significant challenges, including noise, interference, incomplete data, and high deployment costs, which limit their effectiveness and scalability. This paper investigates the potential of Generative AI (GenAI) to overcome these limitations within the IoT ecosystem.
We provide a comprehensive review of state-of-the-art GenAI techniques, focusing on their application to RF sensing problems. By generating high-quality synthetic data, enhancing signal quality, and integrating multi-modal data, GenAI offers robust solutions for RF environment reconstruction, localization, and imaging. Additionally, GenAI's ability to generalize enables IoT devices to adapt to new environments and unseen tasks, improving their efficiency and performance.
The main contributions of this article include a detailed analysis of the challenges in RF sensing, the presentation of innovative GenAI-based solutions, and the proposal of a unified framework for diverse RF sensing tasks. Through case studies, we demonstrate the effectiveness of integrating GenAI models, leading to advanced, scalable, and intelligent IoT systems.
\end{abstract}


\section{Introduction}
\IEEEPARstart{W}{ith} the development of the Internet of Things (IoT), many kinds of wireless sensing signals (e.g., Wi-Fi, Infrared images, visible images, Radio Frequency (RF) signal) are filling our living and working spaces nowadays. Recently, researchers have also utilized RF signals to capture events in the IoT environment (i.e., RF sensing). While RF signals are transmitted, reflected, blocked, and scattered by objects like walls, furniture, vehicles, and human bodies, it is possible to extract useful information, such as position, movement direction, speed, and vital signs of a human subject, from received RF signals. Unlike traditional hardware sensors, RF sensing provides users with low-cost and unobtrusive services. Furthermore, due to the broadcast nature of RF signals, RF sensing can be used not only to monitor multiple subjects, but also to capture changes in the environment over a large area \cite{RFdata}.

Traditional RF sensing methods face several limitations in IoT systems, including noise and interference, which degrade signal quality and lead to inaccurate data interpretation. The presence of other electronic devices and RF sources causes further data loss or corruption. Incomplete data is common in scenarios with limited sensor deployment, and the high costs of deploying and maintaining extensive sensor networks make large-scale implementations expensive. Additionally, unstable environments cause signal weakening and multipath propagation, reducing reliability. These challenges necessitate advanced solutions like Generative AI (GenAI) to enhance the robustness, efficiency, and scalability of IoT systems.

GenAI refer to neural network models designed to generate new data similar to a given dataset, including Generative Adversarial Networks (\textbf{GANs}), Variational Autoencoders (\textbf{VAEs}), Autoregressive Models, flow-Based Models, Diffusion Models (\textbf{DMs}), and Transformer-based Large Language Models (\textbf{LLMs}). These techniques offer significant advantages in data-intensive applications by creating high-quality synthetic data, improving data quality through denoising, and filling in missing values. Generative AI is particularly effective in both cross-modal and multi-modal applications: integrating diverse data types into unified representations for better decision-making and translating information between modalities to enhance robustness. This capability supports innovative IoT applications, smart cities, healthcare, and autonomous systems, showcasing generative AI's transformative potential.

\begin{figure*}[!htbp]
    \centering
    \includegraphics[width=\textwidth]{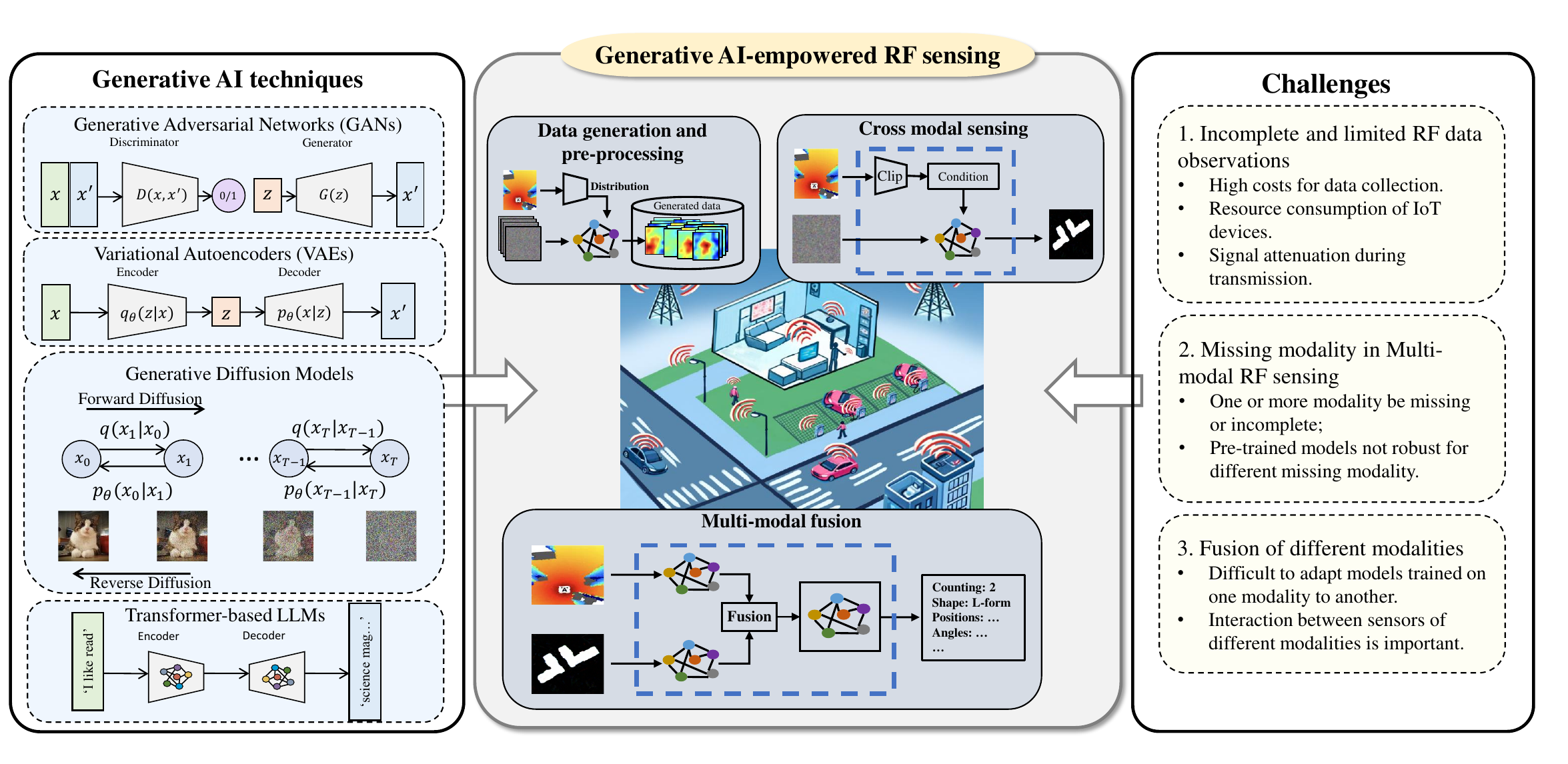}
    \caption{This figure demonstrates an application scenario of RF sensing in IoT systems, highlighting the enhancements provided by generative AI techniques. These enhancements include RF \textbf{data augmentation} to complete missing observations and augment limited real-world data, \textbf{cross-modality generation} to infer missing data from available modalities, and \textbf{multi-modal fusion} to combine information from different data types with RF signals. The figure showcases the use of GANs, VAEs, DMs, and LLMs to address these challenges and improve the overall effectiveness of RF sensing in IoT applications.}
    \label{fig:Figure_synth}
\end{figure*}

GenAI's ability to enhance data quality and integrate various data types makes it ideal for IoT applications, which require universality. With the advent of smarter devices, advanced sensors, and enhanced connectivity technologies like 5G and 6G, IoT systems can greatly benefit from GenAI. It extends conventional deep learning to manage diverse and unforeseen tasks with limited data and resources. GenAI's generalization capability is crucial for IoT devices to adapt to new environments and tasks. Additionally, GenAI's natural language processing enhances multi-modal sensing by integrating text, audio, and visual data, creating more comprehensive and intelligent IoT systems.

As shown in Fig. \ref{fig:Figure_synth}, we discuss the main challenges in RF sensing applications and explore how GenAI can address these issues using unimodal and multi-modal datasets, including reviewing the most relevant works and proposing feasible solutions for the potential use of GenAI. The main contributions of this article can be summarized as follows:
\begin{itemize}
    \item We provide a comprehensive review of state-of-the-art GenAI techniques in RF sensing, covering both uni-modal and multi-modal challenges. We highlight how GenAI models address key RF sensing issues, leveraging their statistical characteristics to model complex data distributions, handle missing data, and enhance signal processing.
    
    \item We discuss the challenges in IoT systems posed by vast data volumes, numerous smart devices, task complexity, and the need for high generalization, particularly in RF sensing. We propose addressing these challenges using GenAI, including LLMs.

    \item We propose a unified RF sensing framework for individual, multiple, and general tasks, and illustrate the effectiveness of integrating GenAI models with a case study.
\end{itemize}

\section{A short review of GenAI techniques}

In this section, a short review of GenAI and their potential in RF sensing applications are provided (see Table \ref{tab:comparison}). GenAI is defined as AI models  which are capable of generating new contents, data, or information from the learned distribution of the trained data. We provide a brief introduction to popular GenAI models and review their application to RF sensing, respectively. 

\textbf{GANs} consist of two neural networks engaging in a zero-sum game: the generator $G$ aims to create new data that matches the statistics of the training dataset, while the discriminato $D$ identifies true and fake data samples. In RF sensing applications, GANs can generate realistic synthetic RF data, enhance signal quality, and enable robust anomaly detection. Their capability to model complex signal environments and facilitate data augmentation contributes to more resilient and adaptive RF sensing systems.

\textbf{VAEs} learn an encoder and decoder to embed data into a continuous latent space, using a parameterized likelihood function that reduces overfitting—making them ideal for large datasets like radio map construction. In RF sensing, VAEs enhance performance by denoising data and reconstructing incomplete information. By understanding the structure and distribution of RF signals, they filter out noise and infer missing parts, creating comprehensive datasets despite partial sensor coverage or data loss. These capabilities improve the robustness, adaptability, and effectiveness of RF sensing systems in various IoT applications.

\textbf{DMs} utilize a Markov chain to add random noise to data (forward diffusion) and then denoise it to generate desired samples (reverse diffusion). This process transforms the data distribution into a simple prior (e.g., Gaussian noise), allowing new data to be generated through denoising. DMs enhance RF sensing in IoT by iteratively generating and denoising data, producing high-quality synthetic RF data and improving signal clarity. They effectively detect anomalies, capture fine variations in RF environments, and robustly reconstruct incomplete RF data, ensuring comprehensive datasets despite partial sensor coverage. These capabilities enhance the precision, reliability, and robustness of RF sensing in IoT applications.

\textbf{LLMs} are pre-trained on vast text datasets to learn contextual dependencies, achieving significant success in applications like question answering, language comprehension, code creation, and reasoning. In IoT systems, LLMs enhance RF sensing by incorporating natural language processing, enabling sensors to interpret and generate human language for smarter device communication. By integrating the natural language modality, LLMs facilitate sophisticated multi-modal data analysis, combining RF data with textual and audio inputs for comprehensive insights. This capability improves anomaly detection, contextual understanding, and decision-making, making IoT systems more intelligent and adaptive.

\definecolor{BlueChalk}{rgb}{0.933,0.882,0.996}
\definecolor{Seashell}{rgb}{0.945,0.945,0.945}
\definecolor{MineShaft}{rgb}{0.2,0.2,0.2}
\begin{table*}
\centering
\caption{The outline of issues encountered in RF sensing for IoT applications, the limitations of traditional AI techniques in addressing these challenges, and the potential of generative AI to enhance performance and efficiency in these contexts.}
\label{tab:comparison}
\begin{tblr}{
  width = \linewidth,
  colspec = {Q[200]Q[362]Q[378]},
  cells = {Thistle, valign = m},
  cells = {BlueChalk, valign = m},
  column{2} = {Seashell},
  cell{1}{2} = {fg=MineShaft},
  cell{1}{3} = {fg=MineShaft},
  cell{2}{2} = {fg=MineShaft},
  hlines,
  vline{2-3} = {-}{},
}
\textbf{  issues} & \textbf{Limitation of traditional deep learning techniques} & \textbf{Potential of Generative AI} \\
{$\cdot$ Incomplete RF data;\\$\cdot$ Limited training data in new environments.} & {$\cdot$ Dependency on large and clean datasets;\\$\cdot$ Lack of Built-in Mechanisms for Missing Data.} & {$\cdot$ Data augmentation and synthesis;\\$\cdot$ Learning data distributions;\\$\cdot$ Generated plausible values from learned data distribution.} \\
{$\cdot$ Strong noise;\\$\cdot$ Interference.} & {$\cdot$ Trained on data with specific noise characteristics;\\$\cdot$ Limited robustness to varying conditions;\\$\cdot$ Tendency to overfit to noise.} & {$\cdot$ Learn the underlying distribution of both signal and noise;\\$\cdot$ Generalize to different noise conditions;\\$\cdot$ Synthesizing training data with various noise and interference scenarios.} \\
{$\cdot$ Missing modalities;\\$\cdot$ Unstable acquisition during test.} & {$\cdot$ Require complete datasets for training and inference;\\$\cdot$ Missing data result in degraded performance.} & {$\cdot$ Integrate information from multiple modalities;\\$\cdot$ Compensate for missing data in one modality using the available data from others.} \\
{$\cdot$ Fuse information in different modality;\\$\cdot$ Low communication efficiency among Internet of Sensors} & {$\cdot$ Static network structures;\\$\cdot$ Difficult to exchange information among sensors;\\$\cdot$ Difficult to adapt to different tasks and dynamic environment.} & {$\cdot$ Provide potential for communication among sensors and intelligent edges;\\$\cdot$ Provide convenience for dynamical environment adaptation.}      
\end{tblr}
\end{table*}

\section{GenAI empowered uni-modal wireless sensing in IoT}

RF sensing utilizes radio frequency signals to detect and interpret physical phenomena, providing essential capabilities for applications like environmental monitoring, health diagnostics, and security systems. Both traditional and modern RF sensing techniques face challenges during data acquisition and transmission, including high costs, significant interference, and bandwidth limitations. Although RF signals are vital in many IoT applications, their sparse nature results from the difficulty and expense of establishing extensive observation points.

\begin{figure}
    \centering
    \includegraphics[width=0.5\textwidth]{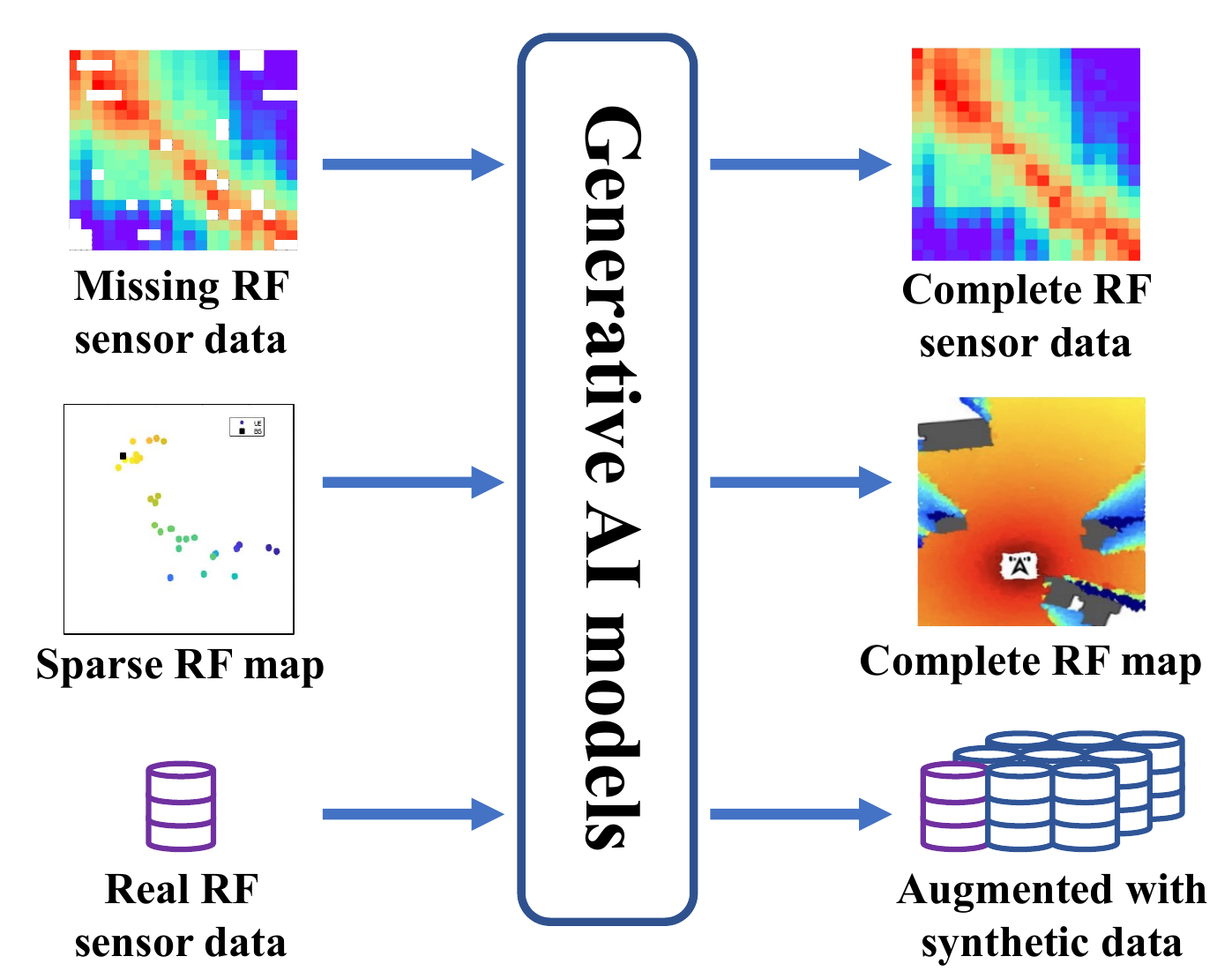}
    \caption{Illustration of GenAI techniques addressing challenges in uni-modal RF sensing for IoT systems through three key examples: completing missing RF sensor data, reconstructing radio maps from extremely sparse observations, and generating synthetic data to enhance model robustness and generalization across diverse environmental conditions.}
    \label{fig:uni-modal illustration}
\end{figure}

To address these challenges, we explore the application of Generative AI techniques in uni-modality RF sensing, inspired by their success in computer vision. As the number of IoT devices increases, efficient signal acquisition and data collection become more complex, leading to issues with synchronization, interference handling, and large data volumes. Consequently, RF signals often suffer from missed observations and sparse structures, hindering effective downstream tasks. Generative AI models, such as GANs, VAEs, and diffusion models, can help reconstruct missing data, infer information for localization, and generate representative samples for downstream tasks (see Fig.~\ref{fig:uni-modal illustration}). The advantages of Generative AI in this context can be categorized into three main aspects.

\subsection{Data imputation} 
A key issue in RF sensing is managing missing Received Signal Strength (RSS) readings, which can reduce accuracy and reliability in applications like localization and environment monitoring. To mitigate this, missing data must be filled in based on observed data, with the correlation between the two being crucial. However, this correlation can vary, and traditional neural networks struggle to accommodate this variability due to their reliance on fixed spatial relationships.
Generative AI models like Transformers are well-suited to address the challenge of missing RSS readings. They learn complex dependencies and the varying impacts of different observations through attention mechanisms that model the significance of each data point. Additionally, GenAI creates embeddings that capture broader context and relationships, improving predictions for missing data. For example, Wang et al. \cite{wang2023radio} utilize the Bidirectional Encoder Representations from Transformers (BERT) model to effectively infer missing values by understanding the contextual relationships within RSS data.
More broadly, the imputation technique using GenAI is particularly valuable for IoT applications with incomplete RF measurements. In smart city infrastructure, for example, environmental factors or structural obstructions can cause intermittent signal loss. GenAI can fill in missing RF data, ensuring consistent monitoring for applications like traffic flow analysis. This approach also benefits intelligent transportation systems, where IoT-based tracking encounters variable RF conditions across different regions. By learning from historical patterns, GenAI can deliver reliable data streams, supporting accurate and uninterrupted monitoring in complex IoT environments.

\subsection{Super-resolution}

While these advancements are promising, there are still significant challenges to address in RF sensing. In particular, in many IoT deployments, the number of sensors is limited due to cost constraints, power requirements, or physical space limitations. This can result in large areas with minimal sensor coverage, leading to extremely sparse data. When there are large missing regions (e.g., 90\% missing), GenAI models are more effective due to their ability to generate data that is not directly present in the input, enabling to infer missing data under large missing rates and complicated loss patterns \cite{zhang2021missing}. 
Among them, DMs have significant potential for super-resolution data generation from sparse latent spaces, especially in complex IoT applications with limited sensor density. In large-scale industrial monitoring, such as for bridges or pipelines, DMs can convert sparse RF input data into detailed condition maps, enhancing structural health monitoring while minimizing sensor deployment. To improve precision, conditioning diffusion models can integrate contextual data, like structural specifications for industrial monitoring or traffic patterns for urban planning, refining the generated data to meet specific needs. Additionally, Large Language Models (LLMs) can be used to create customized guidance for various sensing tasks, such as tailoring prompts for radio maps.

\subsection{Synthetic data generation}
In IoT scenarios, data collection faces logistical, financial, and privacy constraints, along with the complexity of real-world environments. By leveraging AI models to generate high-fidelity synthetic data, it’s possible to augment limited real-world data, enhancing the training and performance of RF signal processing algorithms. Generative AI can improve model robustness by creating synthetic data from learned distributions, introducing controlled variability, and enabling generalization to unseen conditions. Specifically, GenAI can produce data tailored to scenarios with varying environmental conditions, interference patterns, and device configurations, simulating diverse settings like urban, indoor, and rural areas. For instance, Njima et al. \cite{njima2021indoor} proposed a generative adversarial network for RSS data augmentation, successfully generating synthetic RSS data from a small set of real labeled data, which improved localization accuracy.
Synthetic data generation is especially important for applications where real-world data is scarce or costly to obtain. In industrial IoT, synthetic data can simulate harsh environmental conditions and operational variability, allowing models to better predict equipment failures and optimize maintenance schedules. Similarly, in healthcare IoT, where privacy constraints limit data availability, synthetic RF data can enhance model training for indoor localization and patient monitoring systems, supporting accurate and reliable performance across diverse healthcare settings.

\section{GenAI empowered multi-modality RF sensing in IoT}
\begin{figure*}[!htbp]
        \centering
        \includegraphics[width=0.85\textwidth]{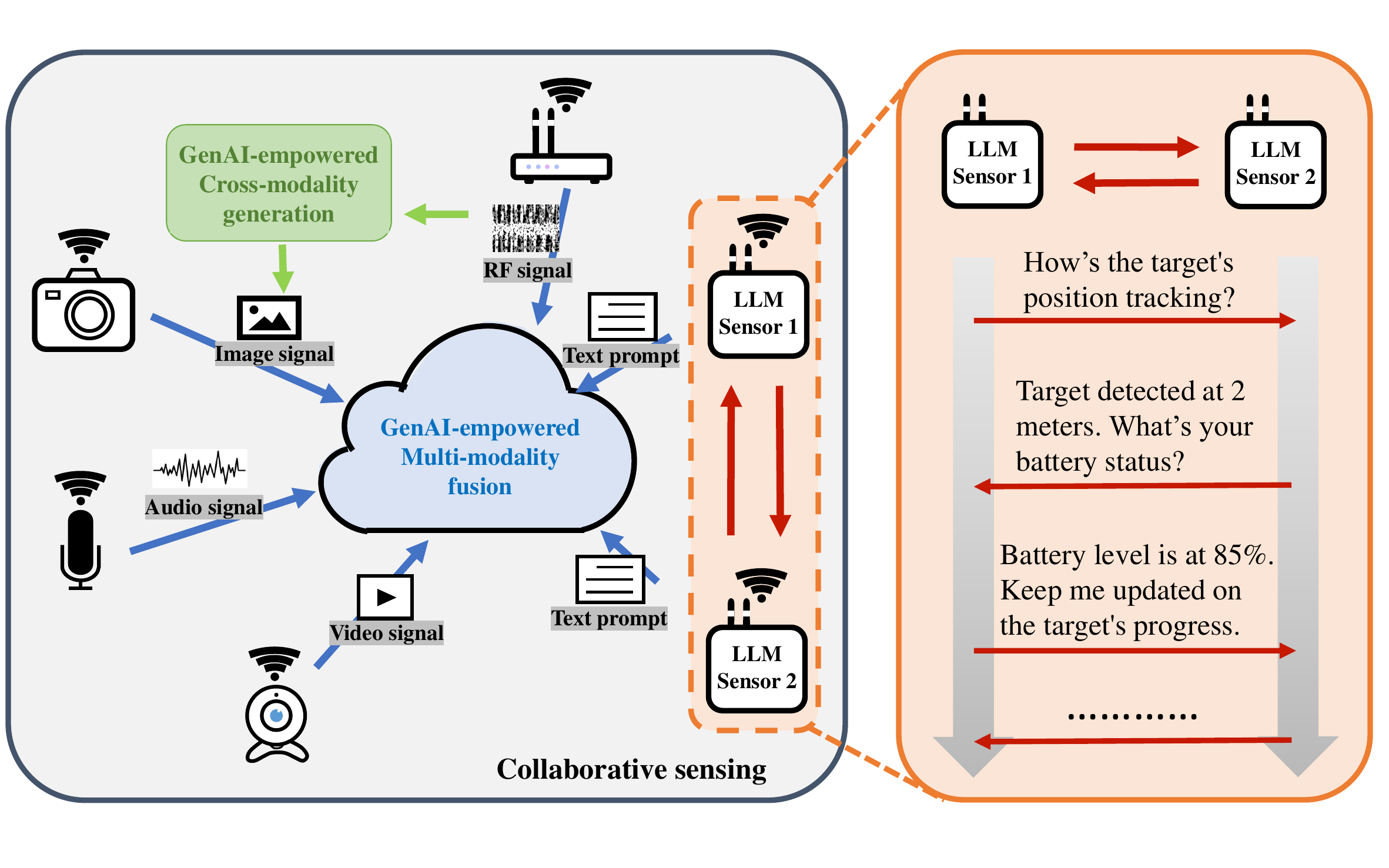}
        \caption{This figure illustrates the scenario of Generative AI-empowered multi-modal RF sensing applications. Three parts of different techniques are presented, namely the \textcolor{darkgreen}{cross-modal generation}, \textcolor{darkblue}{multi-modal fusion} and \textcolor{darkred}{LLM-enhanced interaction among sensors}. The utilization of GenAI for multi-modal sensing makes the Internet of Sensors system more adaptive to dynamic environments and significantly improving overall efficiency.}
        \label{fig:Multi-modal illustration}
    \end{figure*}

In this section, we introduce the multi-modality RF sensing techniques in two parts, namely cross-modality RF sensing and multi-modality fusion for RF sensing (see Fig.~\ref{fig:Multi-modal illustration}). This division is motivated by the capabilities of GenAI to enhance RF sensing performance through distinct approaches. Cross-modality RF sensing leverages GenAI to correlate RF signals with another modality, such as images, to mitigate challenges like sparsity, interference and missing data. Meanwhile, multi-modality fusion combines data from multiple modalities using GenAI to create a comprehensive representation, overcoming the limitations of individual modalities and improving overall sensing performance.

To better highlight the potential of GenAI in each category of techniques, we provide the following definitions:

\begin{definition}
    \textbf{Cross-Modality RF sensing} focuses on learning and utilizing the relationships between different modalities to infer one modality from another. It aims to handle scenarios where one modality may be missing or incomplete, by leveraging information from another modality.
\end{definition}

\begin{definition}
    \textbf{Multi-modality fusion} for RF sensing involves the simultaneous integration of multiple data modalities (e.g., visible images, LiDAR, audios), with RF signals, to create a unified representation that leverages the strengths of each modality and improve overall sensing accuracy and effectiveness. In this approach, it is assumed that multiple modalities are available, and it focuses on combining them to improve performance.
\end{definition}

\subsection{Cross-modal RF sensing with GenAI}

In RF sensing, models are often trained with simulated data or in controlled conditions where signals are clear. However, real-world deployment may involve sparse, noisy, or unavailable RF signals due to environmental constraints or hardware limitations. GenAI-empowered cross-modal RF sensing techniques address these problems through various methodologies.

\textbf{LLMs} offer significant potential for cross-modal learning, enhancing the generalizability of multimodal systems. One strategy involves training universal models on large datasets across various modalities to encode these modalities into embeddings. Multi-modal LLMs generate embeddings that effectively represent underlying information, improving the generalizability of universal models.
Another approach involves converting all modalities into text, creating a model based on text prompts. This method provides a robust system for cross-modal sensing when a modality is missing, as text acts as a unified semantic space, utilizing LLMs' zero-shot prediction capabilities. For example, \cite{tsai2024text} demonstrates how LLMs and foundational models can create a unified semantic space using text, showcasing their ability to perform zero-shot predictions across different modality combinations during testing.

\textbf{VAEs} provide great potential to learn to encode RF signals into a latent space and decode them into another modality. Their probabilistic framework allows for a hierarchical approach to encode data as distributions over latent variables, facilitating diverse outputs in cross-modal tasks. VAEs can learn both joint and modality-specific distributions, enabling effective cross-modality inference, which allows for accurate predictions even with missing or incomplete data by inferring from the shared latent space.
For example, \cite{vasco2020mhvae} introduced a Multimodal Hierarchical Variational Autoencoder (MHVAE), extending the single-modality nature of VAEs to a multimodal hierarchical setting. This approach enables the learning of both modality-specific and joint distributions, demonstrating VAEs' effectiveness in rich multimodal representation and cross-modal generation.

\textbf{DMs} DMs also provide an effective solution for synthesizing cross-modal data by leveraging existing datasets. Their hierarchical statistical structure and iterative denoising process enable high-quality, detailed outputs for tasks like text-to-image generation. DMs adeptly capture and model complex relationships between modalities, making them suitable for advanced cross-modal applications.
For instance, \cite{chen2023rf} proposes a method for synthesizing RF sensing data using cross-modal diffusion models, enhancing the generalization capability of millimeter-wave (mmWave) sensing systems and demonstrating significant improvements in sensing and generalization across two distinct tasks.

Cross-modal generation utilizes GenAI to enhance RF data by learning correlations with more costly or difficult-to-obtain modalities like images and audio. This is crucial in IoT scenarios where RF signals are plentiful but complementary data is scarce. For example, in industrial IoT, RF signals monitor equipment, while inferred thermal data enhances fault detection without the need for expensive thermal sensors. In remote environmental monitoring, RF data provides continuous feedback, supplemented by cross-modal generation to replace costly satellite imagery. This approach allows IoT systems to leverage cost-effective RF data to fill gaps typically covered by more resource-intensive modalities.

\subsection{Multi-modal fusion with GenAI}

The growing demand for real-world applications highlights the limitations of uni-modal sensing, especially for comprehensive environmental understanding. Integrating multi-modal sensing techniques can address this challenge by combining different data types (radio, optical, audio, thermal) with RF signals. This integration provides complementary information, enhancing the accuracy and robustness of sensing tasks. The redundancy from different sensors also improves the reliability and fault tolerance of wireless systems. For instance, in wireless localization, multi-modal techniques enhance positioning accuracy by providing redundant and complementary data, such as visible images, thermal images, and audio, which support and improve conventional uni-modal fingerprinting methods.

\begin{figure*}[!htbp]
\centering
\includegraphics[width=\linewidth]{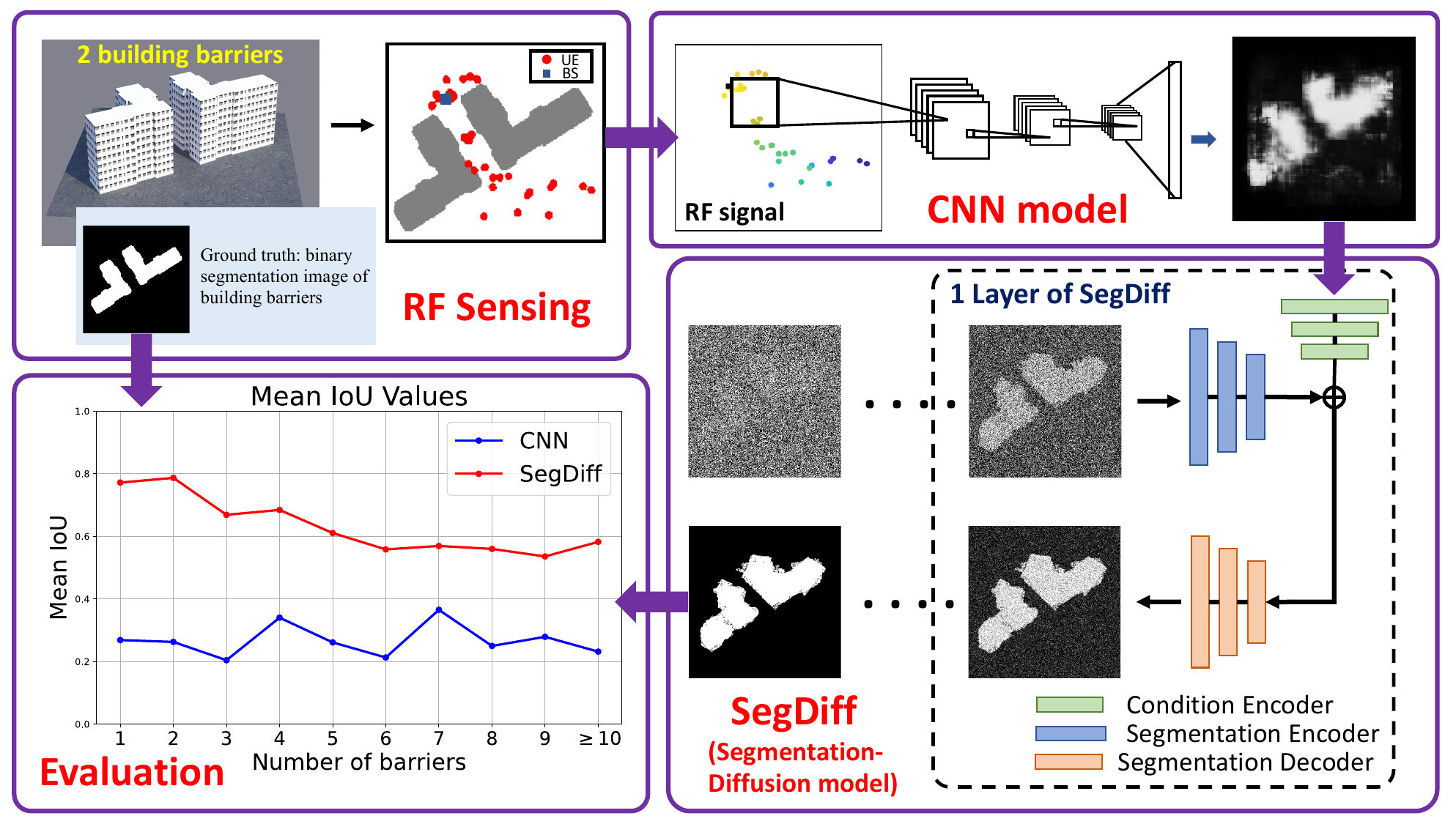}
    \caption{Illustration of the RF sensing technique for building barrier detection based on a Generative AI model utilizing a Segmentation Diffusion model (SegDiff). Initially, a Convolutional Neural Network (CNN) generates a blurred building barrier map from extremely sparse RF signal receive power samples. These blurred barrier maps are then input into the segmentation diffusion model as a condition, which enhances the quality of the detected barrier map by reducing noise and sharpening details. For performance evaluation, the test dataset is divided into 10 classes, with each class representing data containing a specific number of barriers; images with more than 10 barriers are categorized into class 10. The final performance is assessed using the mean Intersection over Union (mean IoU) metric, reflecting the average IoU value across all data within each class. This figure highlights the significant improvements achieved through GenAI, particularly in scenarios with fewer building barriers in the map. The comparison demonstrates the model's ability to detect barrier maps with high accuracy and perceptual quality, showcasing its advantages in RF sensing-based IoT applications.}
\label{fig:diffusive_generation}
\end{figure*}

The advantages of multi-modal sensing come with new challenges, particularly the unique data patterns of each modality, making it hard to adapt models across modalities. Multi-modal fusion techniques, such as the Meta-Transformer \cite{zhang2023meta}, address this issue. A Meta-Transformer uses a data-to-sequence tokenizer to project data into a shared embedding space, a modality-agnostic encoder to process these embeddings, and task-specific heads for downstream predictions. This approach consolidates latent patterns for tasks like wireless positioning, autonomous driving, smart home technologies, and industrial automation.

Integrating GenAI, particularly LLMs, into multi-modality fusion offers transformative potential for RF sensing tasks. GenAI models are highly effective in understanding, generating, and translating complex data across various modalities. LLMs, with their advanced natural language processing, can integrate language as a supplementary modality alongside RF signals, images, and audio. This is particularly useful when sensor data is incomplete or sparse. For example, supplementary sensors can capture and transmit textual descriptions of the environment, such as object locations or specific conditions, with minimal data overhead, reducing costs. LLMs can then fuse this language data with RF signals and other modalities, enriching contextual understanding and filling gaps left by incomplete RF data, thereby enhancing overall sensing capabilities.

\textbf{Enhancing RF Sensing with Language Descriptions}:
1) Consider a scenario where a network of sensors is deployed in an environment for monitoring purposes. While the primary RF sensors capture signals that provide spatial and movement RF data, supplementary sensors record descriptive information about the environment. These descriptions might include details such as ``three people standing near the entrance,'' ``a moving vehicle is approaching from the west,'' or ``there is an obstruction in the path.'' Transmitting these descriptions instead of raw data minimizes the bandwidth requirements and simplifies data collection.
2) Using LLMs, these language descriptions can be fused with the RF signal data to enhance downstream tasks such as localization, object detection, and activity recognition. LLMs can process and integrate the descriptive information, correlating it with the RF data to create a more complete and accurate representation of the monitored environment.
Moreover, the flexibility of LLMs allows them to interpret the descriptive data, making them adapted to scenarios where traditional sensing methods might struggle. For instance, in dynamic environments where RF signals might be highly variable, the additional context provided by language descriptions can significantly improve the reliability and accuracy of the sensing system.
A hybrid End-to-End learning framework for autonomous driving by combining basic driving imitation learning with LLMs based on multi-modality prompt tokens is proposed in \cite{duan2024prompting}, where end-to-end integration of visual and LiDAR sensory are input into learnable multi-modality tokens, and a hybrid setting is explored which use LLMs to help the driving model correct mistakes and complicated scenarios.

\subsection{Case study: GenAI models for RF imaging}
To validate the effectiveness of our Generative AI-empowered RF sensing in IoT systems, we focus on the outdoor environment reconstruction problem using a denoising diffusion model, specifically applying RF sensing to detect barriers in smart city applications.
Outdoor environment reconstruction involves creating digital representations of outdoor spaces, including terrain, buildings, and infrastructure. This process is crucial for urban planning, augmented reality, simultaneous localization and mapping, and industrial network optimization, as it supports infrastructure deployment, signal propagation analysis, and resource allocation.
In our approach, user equipment (UE) and base stations (BS) sample RF signals at extremely sparse locations. These sparsely sampled signals are used for outdoor environment reconstruction, but achieving high-quality barrier detection is challenging due to the limited sample density, which can impact the precision and reliability of the detection process.

GenAI models have been widely used to generate data cross different modalities. For example, causal transformer and U-Net are shown as effective architecture to implement the diffusion prior and diffusion generator.
Inspired by these advancements, we propose an RF sensing-building barrier detection method based on Segmentation Diffusion GenAI model (SegDiff). As illustrated in Figure \ref{fig:diffusive_generation}, the figure highlights the structure of our proposed method, showcasing the advantages of the Generative AI approach over traditional CNN methods in terms of visual quality and detection accuracy, as measured by Intersection over Union (IoU) performance.
Initially, a Convolutional Neural Network (CNN) is trained to generate building barrier maps from extremely sparse receive power samples \cite{huangfu2022wair}, producing a rough representation often characterized by blur and noise. These blurred building barrier map are then fed as the condition of a Segmentation Diffusion model \cite{SegDiff}, which enhances the quality of estimated barrier map by leveraging its learned data distribution to remove noise and sharpen details.

For performance evaluation, we categorize the test RF sensing dataset into 10 classes, with each class representing data containing a specific number of building barriers. Data with more than 10 barriers are classified as class 10. We assess overall detection accuracy using the mean Intersection over Union (mean IoU) metric, which represents the average IoU value across all data within each class. The comparison of barrier detection performance between the CNN method and the SegDiff method demonstrates significant improvements achieved through the use of GenAI. While traditional neural networks struggle to generate clear barrier map from sparse RF sensing data due to their limited capacity to learn exact data distributions, GenAI model outperforms by providing high-quality sensing results.

\section{A foundation model for RF sensing and communication}

\begin{figure*}[!htbp]
\centering
\includegraphics[width=0.9\linewidth]{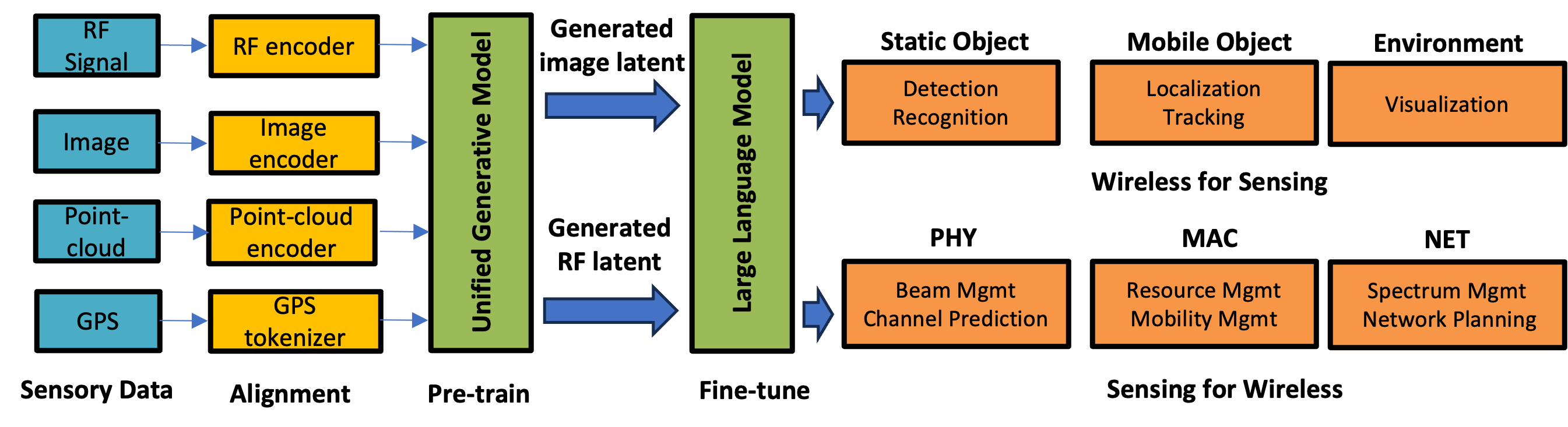}
    \caption{GenAI model for multi-task wireless sensing. A unified GenAI model trained on generating visual or radio images is proposed and shown to integrate an LLM fine-tuned for downstream sensing and communication tasks.}
	\label{fig:sensing}
\end{figure*}

While we have discussed the potential of GenAI models to enhance wireless sensing, an additional research question arises: can we pretrain a foundation GenAI model on sensory data that can be used for different sensing tasks and scenarios? 

Inspired by the Meta-Transformer \cite{zhang2023meta}, a task-agnostic model trained for generating visual data, e.g., 3D environment image, can be used for computer vision tasks such as object detection, visualization and tracking. Similarly, a model trained for generating RF signals, e.g., 4D radio map, can support various wireless communication tasks, such as beam selection, mobility and spectrum management (Fig. \ref{fig:sensing}), via task-specific fine-tuning. Specifically, cross-modality encoders can be trained to encode the raw data on a common latent space. This can be achieved by contrastive learning, following the CLIP \cite{Alec-2021} method to maximize cross-modality similarity between environment image and RF signal. Thereafter, a unified GenAI model is trained on the data embedding to generate the visual or radio environment representation, which can be realized by AR or DM trained on self-supervised learning (i.e. masked or next image/RF token prediction). Finally, the generated visual or RF embedding can be used by a multi-modal LLM to perform multiple downstream tasks in wireless sensing and communications. For example, a visual language model such as \cite{Flamingo} can produce location of a targeted user from the constructed environment image, to complete the localization task. Similarly, a RF-language model can generate policies of optimal beams for tracking a user, to perform the beam management task. After pre-training on large wireless sensing datasets, we anticipate LLMs with embedded visual and radio data can perform multiple tasks in sensing and communications based on the user's prompts. In doing so, we can build a unified model framework for sensing which does not require task specific retraining or fine-tuning, as shown by Fig.~\ref{fig:sensing}. 


\section{Challenges}
In this section, we elaborate on several challenges of using generative AI encountered in RF sensing applications within IoT systems.

\subsection{Model Deployment} 
Deploying generative models in RF sensing is challenging due to high computational demands and the limited resources of IoT devices and wireless sensor networks. Models like Transformers and diffusion models often exceed the processing power and memory of lightweight RF hardware, raising feasibility concerns \cite{SPADE}. Real-time processing complicates matters further by requiring low-latency responses.
Solutions include edge computing and task offloading, where edge servers or cloud nodes manage intensive tasks, allowing IoT devices to focus on lighter functions \cite{Yenduri}. Model compression techniques like quantization and pruning can also reduce model size and computational needs, making deployment on constrained devices more feasible. Additionally, using lightweight, task-specific models for functions such as localization or anomaly detection minimizes device load and enhances adaptability. These strategies collectively help balance performance with the constraints of resource-limited environments.
\subsection{Model Interpretability} 
Generative models’ complex architectures result in "black box" characteristics, limiting interpretability and applicability in high-stakes domains like autonomous navigation. Simplification via distillation or pruning, along with probabilistic outputs to represent prediction confidence, can improve interpretability. Additionally, generative outputs may lack physical validity within RF contexts. Incorporating physics-informed constraints, such as propagation principles, into model structures or loss functions can ensure results align with real-world RF behaviors, enhancing robustness and reliability.
\subsection{Data Requirements} 
Generative AI models in RF sensing demand large, diverse datasets to capture varied environmental conditions, barrier types, and signal fluctuations across IoT settings. Data collection can be labor-intensive, especially when labeled data is scarce or conditions are dynamic, as seen in smart cities and industrial sites. Sparse or incomplete RF data further limits model generalization. To address this, transfer learning, few-shot training, and data augmentation techniques—such as simulation-based methods—can synthetically enrich training datasets, reducing dependency on extensive real-world data.

\subsection{Prior information combination}
A key challenge is effectively utilizing prior information to enhance the accuracy of generated samples from sparse observations. In RF sensing, additional prior information—such as spatial configurations, historical signal patterns, and environmental characteristics—can significantly improve reconstruction accuracy by providing context and reducing ambiguity. To leverage this, it is crucial to integrate prior information seamlessly into the architecture design of generative models. By employing the inherent statistical properties of certain Generative AI models, explainable techniques with statistical frameworks, like physics-informed neural networks, can enhance performance while requiring fewer training resources.

\section{Conclusions}

In this paper, we explore the potential of GenAI for RF sensing to address data acquisition and transmission resource scarcity driven by the increasing demand for IoT systems. Our approach utilizes GenAI for efficient RF signal processing, incorporating uni-modal processing, cross-modal generation, and multi-modal fusion to effectively perform tasks such as radio map reconstruction and localization. We not only present these visions but also demonstrate them through practical case studies, discussing the challenges and opportunities, and proposing GenAI-based solutions as promising avenues for future developments in RF sensing.
We foresee GenAI significantly benefiting various IoT applications: enhancing smart city operations by accurately predicting vehicle movements and reducing congestion through real-time data analysis; improving wearable IoT devices for better patient outcomes and reduced healthcare costs; and advancing predictive maintenance in industrial IoT by analyzing sensor data to predict equipment failures. GenAI has the potential to revolutionize our interactions with interconnected systems, leading to smarter, more efficient, and responsive IoT environments.

\newpage

\end{document}